\documentclass[prd,twocolumn, nofootinbib]{revtex4}
\usepackage{amsmath}
\usepackage{graphicx}
\usepackage{dcolumn}
\usepackage{bm}
\usepackage{amssymb}
\usepackage{latexsym}
\usepackage{subfigure}
\usepackage{epsfig}

\newcommand{\be}{\begin{equation}}
\newcommand{\ee}{\end{equation}}
\newcommand{\ba}{\begin{eqnarray}}
\newcommand{\ea}{\end{eqnarray}}

\def\der{{\rm d}}

\def \rsdv{r_{\rm d}/D_{\rm V}}
\def \rsda{r_{\rm d}/D_{\rm A}}
\def \rsdh{r_{\rm d}/D_{\rm H}}
\def \dars{D_{\rm A}/r_{\rm d}}
\def \dhrs{D_{\rm H}/r_{\rm d}}

\begin{document}
\title{The consistency test on the cosmic evolution}
\author{Yan Gong$^{1}$}\email{gongyan@bao.ac.cn}\author{Yin-Zhe Ma$^{2}$}\author{Shuang-Nan Zhang$^{3,1}$}\author{Xuelei Chen$^{1,4}$}
\affiliation{$^{1}$ National Astronomical Observatories,
Chinese Academy of Sciences, Beijing, 100012, China \\ $^{2}$Jodrell Bank Centre for Astrophysics, School of Physics and Astronomy, University of Manchester, Manchester, M13 9PL, UK \\ $^{3}$Key Laboratory of Particle Astrophysics, Institute of High Energy Physics, Chinese Academy of Sciences, P.O.Box 918-3, Beijing 100049, China\\ $^4$ Center of High Energy Physics, Peking University, Beijing 100871, China}

\begin{abstract}
We propose a new and robust method to test the consistency of the cosmic evolution given by a cosmological model. It is realized by comparing the combined quantity $r_{\rm d}^{\rm CMB}/D_{\rm V}^{\rm SN}$, which is derived from the comoving sound horizon $r_{\rm d}$ from cosmic microwave background (CMB) measurements and the effective distance $D_{\rm V}$ derived from low-redshift Type-Ia supernovae (SNe Ia) data, with direct and independent $r_{\rm d}/D_{\rm V}$ obtained by baryon acoustic oscillation (BAO) measurements at median redshifts. We apply this test method for the $\rm \Lambda$CDM and $w$CDM models, and investigate the consistency of the derived value of $\rsdv$ from $Planck$ 2015 and the SN Ia data sets of Union2.1 and JLA ($z<1.5$), and the $\rsdv$ directly given by BAO data from six-degree-field galaxy survey (6dFGS), Sloan Digital Sky Survey Data Release 7 Main Galaxy Survey (SDSS-DR7 MGS), DR11 of SDSS-III, WiggleZ and Ly$\alpha$ forecast surveys from Baryon Oscillation Spectroscopic Data (BOSS) DR-11 over $0.1<z<2.36$. We find that $r_{\rm d}^{\rm CMB}/D_{\rm V}^{\rm SN}$ for both non-flat $\Lambda$CDM and flat $w$CDM models with Union2.1 and JLA data are well consistent with the BAO and CMB measurements within $1\sigma$ CL. Future surveys will further tight up the constraints significantly, and provide stronger test on the consistency.
\end{abstract}


\maketitle

\email{E-mail:gongyan@bao.ac.cn}

\section{Introduction}
\label{sec:intro}

In the last two decades, cosmologists have been making great efforts for establishing standard cosmological model to describe the contents and evolution of the Universe. With precise measurements of the cosmic microwave background from the {\it Wilkinson Microwave Anisotropy Probe} ({\it WMAP})~\cite{Hinshaw13,Bennett13} and {\it Planck}~\cite{Planck13-16,Planck15-13} satellites, the cosmological parameters have been measured in a higher and higher precision, making it possible to test whether the standard $\Lambda$ cold dark matter ($\Lambda$CDM) model can describe the cosmic evolution throughout the history of the Universe. In fact, after {\it Planck} has published its 2013 results~\cite{Planck13-16}, it was found that the previously measured value of Hubble constant $H_{0}$ through $600$ Cepheid variables~\cite{Riess11} is higher than the {\it Planck} measured value by $3\sigma$ confidence level, although later it is shown that by correcting the NGC 4258 distance one can obtain a lower value of $H_{0}$ which is compatible with {\it Planck} 2013 results~\cite{Efstathiou14}. In addition, it is shown that the {\it Planck} constrained $\Omega_{\rm M}$--$\sigma_{8}$ parameter plane is in tension with CFHTLenS data~\cite{Planck15-13}, thermal Sunyaev-Zeldovich effect~\cite{Planck15-22}, and statistics of cluster number counts~\cite{Planck15-24}. These interesting tensions between cosmological data sets drive us to consider more, and robust test on $\Lambda$CDM model at different periods of cosmic evolution.

A useful and interesting data set of cosmic distance estimator is the baryon acoustic oscillation (BAO) data from galaxy surveys~\cite{Anderson14,Beutler11,Blake12,Delubac15}. The measurement of BAO scale is normally written as $r_{\rm d}/D_{\rm V}(z)$ (or $D_{\rm V}(z)/r_{\rm d}$). Here $r_{\rm d}$ is the comoving sound horizon at the end of the baryon drag epoch, which is completely determined by the physics in the early Universe at redshift $z \gtrsim 1100$. In this regime, the observation of the cosmic microwave background radiation can faithfully reflect the physics that prevails the Universe. After the photon decoupled, the baryons began to fall into the initial gravitational potential provided by dark matter, so that the galaxies formed in the potential well at the late-time cosmic evolution. Therefore, the gravitational clustering scale of the BAO seen in galaxy redshift surveys ($\sim 105\,h^{-1}$Mpc) is fixed and determined by $r_{\rm d}$ and does not depend on late-time cosmic evolution. On the other hand, the denominator of the BAO measurement, $D_{\rm V}(z)$, is determined by the late-time cosmic expansion, i.e. angular diameter distance and Hubble parameter, which can be strongly affected by dark energy or modified gravity effect.

Therefore we would like to perform a test of cosmic evolution in a different perspective than Refs.~\cite{Planck13-16,Planck15-13}. If a cosmology model is
the true model of the Universe, it should be able to fit observations throughout all redshift ranges. Therefore, since Type-Ia supernova  (SN Ia) data is mainly obtained at low redshift ($z<1.5$), and CMB data at high redshift ($z \simeq 1100$), we will use CMB to determine the $r_{\rm d}$ only, and use SN Ia data to determine $D_{\rm V}(z)$. Then we compare the combined quantity $r_{\rm d}^{\rm CMB}/D_{\rm V}^{\rm SN}$ with the direct measurements of $r_{\rm d}/D_{V}(z)$ from BAO surveys at median redshifts out to $z\simeq 2.4$. In this way, we calculate the lever arm at two ends (CMB and SN Ia) and compare their prediction in the intermediate redshift regime (BAO). We can then obtain a robust test on cosmic evolution by independently considering different data sets.

This paper is organized as follows: in Section~\ref{sec:obs_data}, we discuss the observational data we use, and show the details of the calculation in the comparison for each data set, especially for the SN Ia data; in Section~\ref{sec:compare}, we talk about the comparison results of $\rsdv$ and the other quantities used in the comparison for the $\rm \Lambda$CDM and $w$CDM models with all three kinds of measurements; the summary and conclusion will be presented in the last section.

\section{Observational data}
\label{sec:obs_data}

In this section, we discuss the data we use in the comparison, including the SN Ia data from Union2.1 and JLA, the BAO data from galaxy and Ly$\alpha$ forest surveys and the sound horizon data from the results of $Planck$ 2015. We also show the details of the derivation of the volume weighted effective distance $D_{\rm V}$ from low-redshift SNe Ia, $\rsdv$ from mid-redshift BAO and the sound horizon $r_{\rm d}$ from high-redshift CMB measurements.

\subsection{SN Ia data}

\begin{figure*}
\centerline{\includegraphics[width=3.3in]{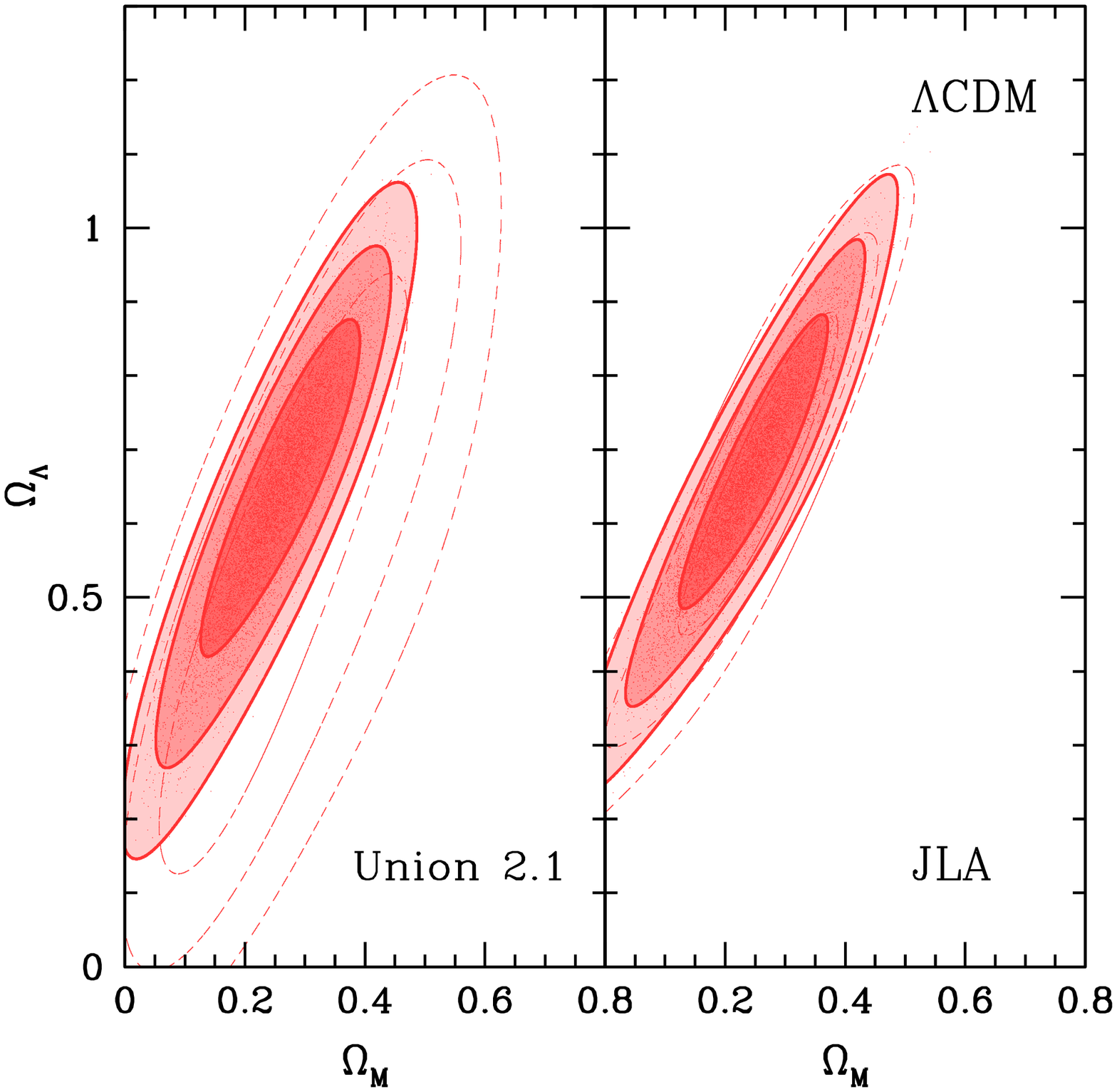}
\includegraphics[width=3.3in]{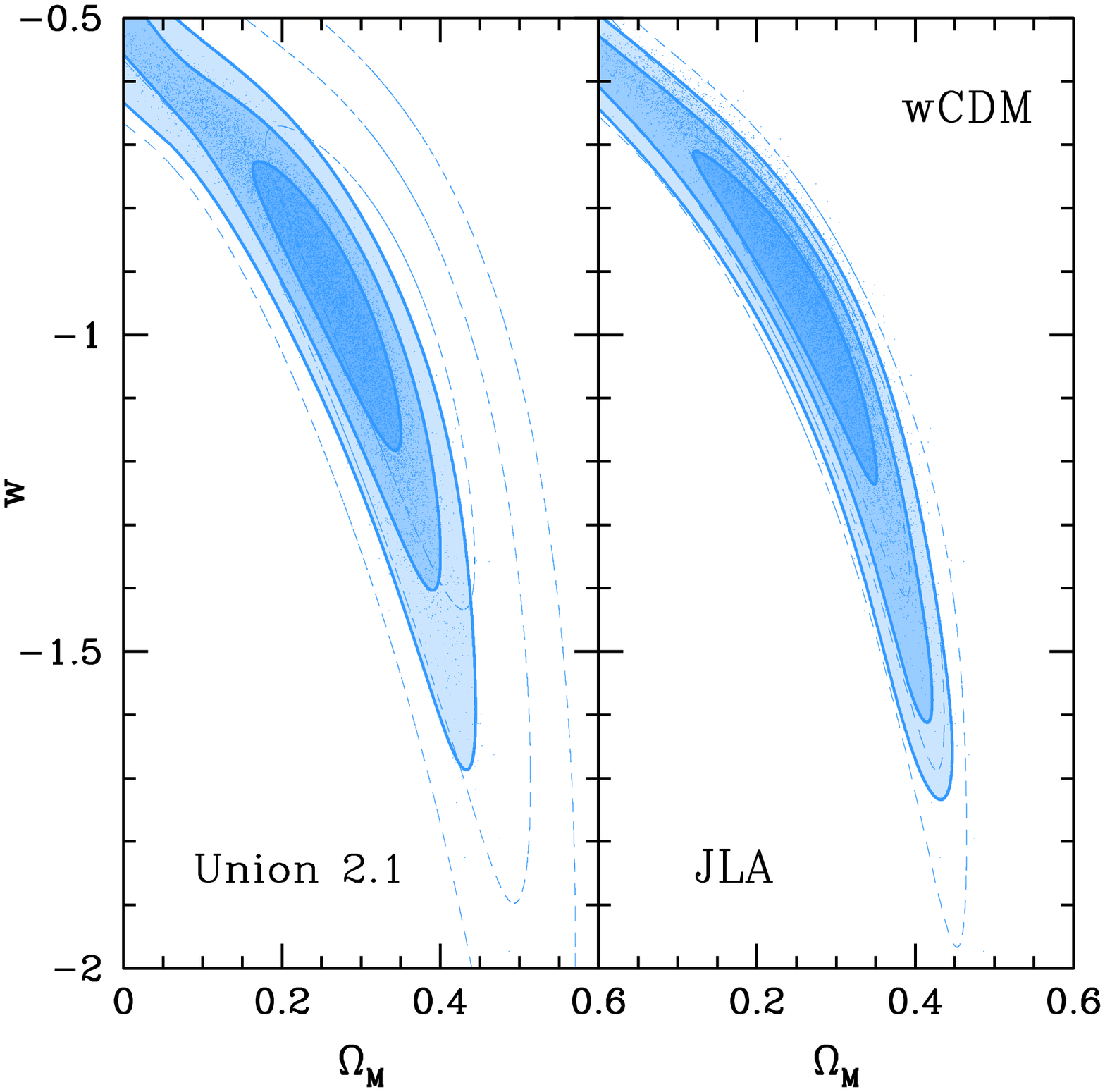}}
\caption{\label{fig:m0l0_m0w} The joint constraints on $\Omega_{\rm M}$ vs. $\Omega_{\rm \Lambda}$ for non-flat $\rm \Lambda$CDM model ({\it left}), and $\Omega_{\rm M}$ vs. $w$ for flat $w$CDM model ({\it right}). The constraint results from both of Union2.1 and JLA data sets are shown here. The solid filled contours and dashed contours are the results without and with systematic errors, respectively. The dots are the MCMC chain points used to illustrate the contour maps for the case without systematic errors.}
\end{figure*}

\begin{figure}[t]
\centerline{\includegraphics[width=3.3in]{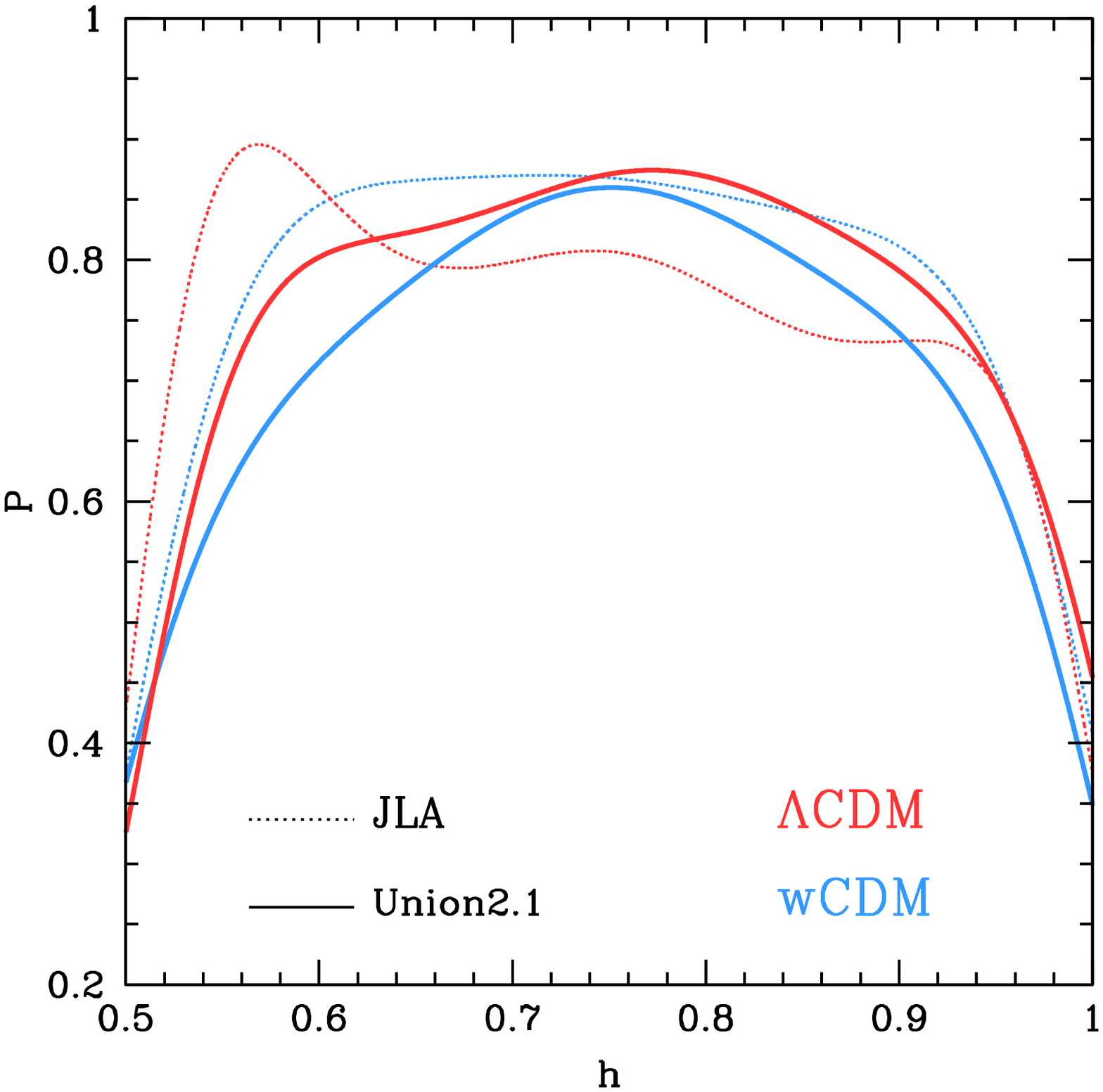}}
\caption{\label{fig:h} The marginalized probability distribution function (PDF) of $h$ for non-flat $\rm \Lambda$CDM and flat $w$CDM models with systematic errors. The solid and dotted curves are for Union2.1 and JLA data sets, respectively. As can be seen, the constraints are loose for both $\rm \Lambda$CDM and $w$CDM models, and the PDFs are relatively flat over the prior range. The results are similar for the cases without systematic errors.}
\end{figure}

There have been a long history of using SN Ia data to constrain the evolution of the Universe. However, in recent years, as the constraints from CMB measurement have been greatly improved, it is found that some SN Ia samples provide $\sim 2\sigma$ inconsistent results than BAO and CMB measurement. For instance, in~\cite{Planck13-16}, it is found that by using the SNLS combined Type-Ia supernovae samples, the joint constraints on $w_{0}$--$w_{a}$ parameters prefer a phantom dark energy ($w<-1$) which is inconsistent with the {\it Planck}+BAO measurement at $2\sigma$ CL. Same problem is also found in~\cite{Shafer14} and \cite{Karpenka15}. Interestingly, by re-calibrating the light curve fitting parameters with BAO data set, Ref.~\cite{Li14} finds that the results of the SN constraints shift back to be consistent with {\it Planck} measurement of the CMB. Thus, there is a possibility that the light curve parameters, or the distance-modulus relation in SNLS samples have some systematic bias so that the results deviate from the true values at $2\sigma$ CL. To avoid or reduce this possible bias, in our approach we use the SN Ia data from Union2.1 catalog\footnote{\tt http://supernova.lbl.gov/union} \cite{Suzuki2012,Hazra15} and the Joint Light-curve Analysis (JLA) SN Ia data\footnote{\tt{http://supernovae.in2p3.fr/sdss\_snls\_jla/ReadMe.html}} \cite{Betoule14}. These two data sets contain the largest SN Ia samples at present, and could provide stringent constraints on the cosmological parameters and the quantities we are interested in.

The Union2.1 sample contains 580 SNe Ia which are calibrated by SALT2 light-curve fitter \cite{Guy2007}, and its redshift range covers $0.015<z<1.414$ \cite{Suzuki2012}. Following \cite{Suzuki2012}, the observed distance modulus of Union2.1 data set is given by
\be
\mu_{B}^{\rm Union} = m_{B}^{\rm max} + \alpha\cdot x_1 - \beta\cdot c + \delta\cdot P - M_B,
\ee 
where $m_B^{\rm max}$, $x_1$ and $c$ are the three parameters of light curve that are fitted by SALT2 \cite{Guy2007}. The $m_B^{\rm max}$ is the rest-frame $B$-band peak magnitude, $x_1$ and $c$ are the light-curve shape and color parameters respectively. The parameter $P$ denotes the possibility that SNe Ia belong to the host galaxy with mass less than $10^{10}$ $m_{\odot}$, which takes account of the correlation between the SN Ia luminosity and host galaxy mass. The $M_B$ is the absolute $B$-band SN Ia magnitude with $x_1=0$, $c=0$ and $P=0$. The $\alpha$, $\beta$, $\delta$ and $M_B$ are the nuisance parameters which need to be fitted with the cosmological parameters.

The JLA data set includes several low-redshift samples ($z < 0.1$), three season samples from the SDSS-II ($0.05 < z < 0.4$), and three-year data from SNLS ($0.2 < z < 1$). In total, it consists of 740 spectroscopically confirmed SNe Ia with high-quality light curves~\cite{Betoule14}. For the JLA data, the observed distance modulus is 
\be
\mu_{B}^{\rm JLA} = m_{B}^{\rm max} + \alpha\cdot x_1 - \beta\cdot c - (M_B+\Delta_M),
\ee
and $\Delta_M=0$ for $M_{\rm gal}<10^{10}$ $M_{\odot}$, where $M_{\rm gal}$ is the mass of host galaxy. The $\Delta_M$ denotes the correction of the absolute magnitude $M_B$ for $M_{\rm gal}\geq 10^{10}$ $M_{\odot}$, and it is a nuisance parameter that would be fitted with the cosmological parameters.

On the other hand, the theoretical distance modulus can be estimated by
\be
\mu^{\rm th}(z) = 5\,{\rm log_{10}}D_{\rm L}(z) + 25.
\ee
Here $D_{\rm L}(z)=(1+z)\,D_{\rm C}(z)$ (in Mpc) is the luminosity distance from redshift $z$, where $D_{\rm C}$ is the comoving distance and is given by 
\be
D_{\rm C}(z) = |\Omega_{\rm k}|^{-1/2} {\rm sinn}\left(|\Omega_{\rm k}|^{1/2} \int_0^z \frac{c\,{\rm d}z'}{H(z')}\right),
\ee
where $\Omega_{\rm k}$ is the cosmic curvature parameter, and ${\rm sinn}(x)={\rm sinh}(x)$, $x$, sin$(x)$ for open, flat and closed cosmic geometries, respectively. The $H(z) = H_0\Omega(z)^{1/2}$ is the Hubble parameter where $H_0=100\,h$ $\rm km\,s^{-1} Mpc^{-1}$, and $\Omega(z)$ is expressed by
\be
\Omega(z) = \Omega_{\rm M}(1+z)^3 + \Omega_{\rm DE}(1+z)^{3(1+w)} + \Omega_{\rm k}(1+z)^2.
\ee
Here $\Omega_{\rm M}+\Omega_{\rm DE}+\Omega_{\rm k}=1$, and $\Omega_{\rm M}$ and $\Omega_{\rm DE}$ are the fractional energy densities of the matter and dark energy, respectively. The $w$ is the static equation of state of dark energy. We have $\Omega_{\rm DE}=\Omega_{\rm \Lambda}$ when $w=-1$ for non-flat $\rm \Lambda$CDM model, and $\Omega_{\rm k}=0$ for flat $w$CDM model.

Therefore, the cosmological parameters are $\Omega_{\rm M}$, $\Omega_{\rm \Lambda}$ and $h$ in non-flat $\rm \Lambda$CDM model, and $\Omega_{\rm M}$, $w$ and $h$ in flat $w$CDM model. Note that the Hubble constant $H_0$ (or equivalently $h$) has large degeneracy with the other cosmological parameters as shown in the $H(z)$ definition, and the SN Ia data actually cannot provide good constraint on it. However, to avoid introducing priors and affecting the fitting results, we set $H_0$ (or $h$) as a free parameter instead of fixing it to be 70 $\rm km\,s^{-1} Mpc^{-1}$ \cite{Suzuki2012,Betoule14}. 

We adopt the $\chi^2$ distribution to estimate the likelihood function $\mathcal{L}\propto {\rm exp}(-\chi^2/2)$, and we have
\be \label{eq:chi2}
\chi^2 = \sum^{N_{\rm d}}_{i=1} \frac{\left[\mu^{\rm obs}_B({\bf p}_l)-\mu^{\rm th}(z,{\bf p}_c)\right]^2}{\sigma^2},
\ee
where $N_{\rm d}$ is the number of SN Ia data. The ${\bf p}_c$ is the cosmological parameter sets, and ${\bf p}_c=(\Omega_{\rm M}, \Omega_{\rm \Lambda}, h)$ and $(\Omega_{\rm M}, w, h)$ for non-flat $\rm \Lambda$CDM and flat $w$CDM model, respectively. The ${\bf p}_l$ denotes the nuisance parameter sets of the light curve, and we have ${\bf p}_l=(\alpha, \beta, \delta, M_B)$ and $(\alpha, \beta, \Delta_M, M_B)$ for Union2.1 and JLA, respectively. The $\sigma$ is the error for each SN Ia measurement, which is given by $\sigma^2=\sigma_{\rm lc}^2+\sigma^2_{\rm ext}+\sigma^2_{\rm sys}$. The $\sigma_{\rm lc}$ is the error of light-curve parameters, and $\sigma_{\rm ext}$ includes the uncertainties of host galaxy peculiar velocity \footnote{We assume 300 and 150 km s$^{-1}$ for Union2.1 and JLA data, respectively \citep{Suzuki2012,Betoule14}.} and gravitational lensing effect \footnote{We take $\sigma_{\rm lens}=0.093z$ and 0.055$z$ for Union2.1 and JLA data, respectively \citep{Suzuki2012,Betoule14}.}. Here we also include  systematic errors $\sigma_{\rm sys}$ in our estimation. Note that $\sigma_{\rm sys}$ is obtained by different methods for Union2.1 and JLA data, which is derived by setting the reduced $\chi^2$ to be unity for each SN Ia subsample in Union2.1 and by REML method in JLA \citep{Suzuki2012,Betoule14,Harville1977}. The full covariance matrix between distance modulus are used for Union2.1 data with systematic errors, and we adopt the covariance matrix $\bf C$ in Eq. (\ref{eq:chi2}) instead of $\sigma^2$ in this case.

We use the Markov Chain Monte Carlo (MCMC) to constrain the free parameters (nuisance parameters of the light curve and cosmological parameters). The Metropolis-Hastings algorithm is employed to determine the probability of accepting the new chain points \cite{Metropolis1953,Hastings1970}, and the proposal density matrix is evaluated by a Gaussian sampler with adaptive step size \cite{Doran2004}. We assume uniform prior distribution for all free parameters, and their ranges in the MCMC fitting process are set to be as follow: $\Omega_{\rm M}\in(0, 1)$, $\Omega_{\rm \Lambda}\in(0,2)$, $w\in(-3,0)$, $h\in(0.5, 1)$, $\alpha\in(0, 3)$, $\beta\in(0, 5)$, $M_B\in(-21, -17)$, $\delta\in(-0.3, 0.3)$ and $\Delta_M\in(-0.3, 0.1)$. These ranges are chosen by our MCMC pre-runs and the relevant results from \citep{Suzuki2012} and \citep{Betoule14}. We find the fitting results are not very sensitive to the widths of these ranges. We perform sixteen parallel chains and get about $10^5$ points for each chain after the convergence is reached \cite{Gelman1992}. After performing burn-in process and thinning the chains, we merge all chains together and obtain about 10,000 points to illustrate the probability distribution function of the free parameters \cite{Gong2007}.

In Fig.~\ref{fig:m0l0_m0w}, we show the MCMC constraint results of $\Omega_{\rm M}$ vs. $\Omega_{\rm \Lambda}$ and $\Omega_{\rm M}$ vs. $w$ for non-flat $\rm \Lambda$CDM and flat $w$CDM models. We find Union2.1 data give similar constraints to JLA data without systematic errors, and the best-fit values of the parameters are well consistent in 1-$\sigma$ for these two data sets. When considering the systematic errors, the deviations of the best-fit values of $\Omega_{\rm M}$, $\Omega_{\rm \Lambda}$ and $w$ between Union2.1 and JLA become larger but are still within 1-$\sigma$ CL. We also notice that the contours with systematic errors are comparable to the case without systematic errors for JLA data. This could be due to the fact that there is no off-diagonal components for the covariance matrix of systematic errors between different SNe in JLA data, and just an ${\sigma}_{\rm sys}$ ($\simeq 0.1$) is added to each SN subsample as the systematic error \citep{Betoule14}.

We also investigate the constraints on $h$ (or equivalently $H_0$) as shown in Fig.~\ref{fig:h}. Although there are peak features in the PDFs of $h$, we find it is not well constrained by SN Ia data only as expected. The PDFs are basically flat over the range from 0.5 to 1 for both Union2.1 and JLA data. Here we emphasize that this flexibility of the constraints on $h$ can provide unbiased constraints on the other cosmological parameters without introducing strong prior effects by fixing $h$ or the Hubble constant $H_0$. Hence it could give reliable constraint results of $\Omega_{\rm M}$, $\Omega_{\rm \Lambda}$ and $w$  from SN Ia data only as well as the effective volume distance $D_V$.

\begin{table*}
\begin{center}
\caption{The best-fit values and 1-$\sigma$ errors of the cosmological and light-curve parameters from the MCMC fitting for Union2.1 and JLA data.} \label{tab:fitting}
\vspace{4mm}
\begin{tabular}{ l | l | l | l | l | l | l | | | | | | | | | l | l | l | l | l | l | }
\hline\hline
\multicolumn{1}{c|} {Data} & \multicolumn{4}{|c|}{Union2.1}  & \multicolumn{4}{|c}{JLA} \\
\hline 
\multicolumn{1}{c|} {Model} & \multicolumn{2}{|c|} {non-flat $\Lambda$CDM} & \multicolumn{2}{|c|}{flat $w$CDM} & \multicolumn{2}{|c|} {non-flat $\Lambda$CDM} & \multicolumn{2}{|c}{flat $w$CDM} \\
 \hline
\multicolumn{1}{c|} {Error} & \multicolumn{1}{|c|}{no sys} & \multicolumn{1}{|c|}{sys} &\multicolumn{1}{|c|}{no sys} & \multicolumn{1}{|c|}{sys} & \multicolumn{1}{|c|}{no sys} & \multicolumn{1}{|c|}{sys} &\multicolumn{1}{|c|}{no sys} & \multicolumn{1}{|c}{sys} \\
\hline
\multicolumn{1}{c|} {$\Omega_{\rm M}$ } & \multicolumn{1}{|c|} {$0.26\pm 0.08$} & \multicolumn{1}{|c|}{$0.28^{+0.12}_{-0.13}$} & \multicolumn{1}{|c|} {$0.28^{+0.07}_{-0.10}$} & \multicolumn{1}{|c|}{$0.35^{+0.10}_{-0.16}$}  & \multicolumn{1}{|c|} {$0.24\pm 0.08$} & \multicolumn{1}{|c|}{$0.25\pm0.09$} & \multicolumn{1}{|c|} {$0.28^{+0.07}_{-0.10}$} & \multicolumn{1}{|c}{$0.30^{+0.07}_{-0.10}$} \\

\multicolumn{1}{c|} {$\Omega_{\rm \Lambda}$} & \multicolumn{1}{|c|} {$0.65\pm 0.13$} & \multicolumn{1}{|c|}{$0.56^{+0.21}_{-0.24}$} & \multicolumn{1}{|c|} {$-$} & \multicolumn{1}{|c|}{$-$}  & \multicolumn{1}{|c|} {$0.69\pm0.12$} & \multicolumn{1}{|c|}{$0.66^{+0.15}_{-0.14}$} & \multicolumn{1}{|c|} {$-$} & \multicolumn{1}{|c}{$-$} \\

\multicolumn{1}{c|} {$w$} & \multicolumn{1}{|c|} {$-$} & \multicolumn{1}{|c|}{$-$} & \multicolumn{1}{|c|} {$-0.90^{+0.24}_{-0.19}$} & \multicolumn{1}{|c|}{$-0.74^{+0.18}_{-0.40}$}  & \multicolumn{1}{|c|} {$-$} & \multicolumn{1}{|c|}{$-$} & \multicolumn{1}{|c|} {$-0.92^{+0.20}_{-0.22}$} & \multicolumn{1}{|c}{$-0.94^{+0.27}_{-0.24}$} \\

\multicolumn{1}{c|} {$h$} & \multicolumn{1}{|c|} {$0.92^{+0.07}_{-0.38}$} & \multicolumn{1}{|c|}{$0.77^{+0.21}_{-0.25}$} & \multicolumn{1}{|c|} {$0.91^{+0.08}_{-0.38}$} & \multicolumn{1}{|c|}{$0.75\pm0.22$}  & \multicolumn{1}{|c|} {$0.88^{+0.10}_{-0.37}$} & \multicolumn{1}{|c|}{$0.57^{+0.41}_{-0.06}$} & \multicolumn{1}{|c|} {$0.56^{+0.41}_{-0.05}$} & \multicolumn{1}{|c}{$0.72^{+0.27}_{-0.20}$} \\

\multicolumn{1}{c|} {$\alpha$} & \multicolumn{1}{|c|} {$0.11\pm 0.01$} & \multicolumn{1}{|c|}{$0.10\pm 0.01$} & \multicolumn{1}{|c|} {$0.11\pm 0.01$} & \multicolumn{1}{|c|}{$0.10\pm0.01$}  & \multicolumn{1}{|c|} {$0.13\pm 0.01$} & \multicolumn{1}{|c|}{$0.13\pm0.01$} & \multicolumn{1}{|c|} {$0.13\pm 0.01$} & \multicolumn{1}{|c}{$0.13\pm0.01$} \\

\multicolumn{1}{c|} {$\beta$} & \multicolumn{1}{|c|} {$2.31\pm{0.05}$} & \multicolumn{1}{|c|}{$2.29\pm0.06$} & \multicolumn{1}{|c|} {$2.30^{+0.06}_{-0.05}$} & \multicolumn{1}{|c|}{$2.29^{+0.06}_{-0.05}$}  & \multicolumn{1}{|c|} {$3.13\pm0.08$} & \multicolumn{1}{|c|}{$2.96\pm 0.09$} & \multicolumn{1}{|c|} {$3.12\pm0.08$} & \multicolumn{1}{|c}{$2.96\pm0.09$} \\

\multicolumn{1}{c|} {$M_B$} & \multicolumn{1}{|c|} {$-18.70^{+0.15}_{-0.66}$} & \multicolumn{1}{|c|}{$-19.17^{+1.99}_{-1.63}$} & \multicolumn{1}{|c|} {$-18.70^{+0.15}_{-0.72}$} & \multicolumn{1}{|c|}{$-19.62^{+2.46}_{-1.20}$}  & \multicolumn{1}{|c|} {$-18.51^{+0.17}_{-0.68}$} & \multicolumn{1}{|c|}{$-18.50^{+0.16}_{-1.17}$} & \multicolumn{1}{|c|} {$-18.49^{+0.15}_{-1.24}$} & \multicolumn{1}{|c}{$-18.53^{+0.18}_{-0.99}$} \\

\multicolumn{1}{c|} {$\delta$} & \multicolumn{1}{|c|} {$-0.03\pm 0.03$} & \multicolumn{1}{|c|}{$-0.14^{+0.43}_{-0.14}$} & \multicolumn{1}{|c|} {$-0.03\pm0.03$} & \multicolumn{1}{|c|}{$-0.03^{+0.31}_{-0.24}$}  & \multicolumn{1}{|c|} {$-$} & \multicolumn{1}{|c|}{$-$} & \multicolumn{1}{|c|} {$-$} & \multicolumn{1}{|c}{$-$} \\

\multicolumn{1}{c|} {$\Delta_M$} & \multicolumn{1}{|c|} {$-$} & \multicolumn{1}{|c|}{$-$} & \multicolumn{1}{|c|} {$-$} & \multicolumn{1}{|c|}{$-$}  & \multicolumn{1}{|c|} {$0.04^{+0.06}_{-0.34}$} & \multicolumn{1}{|c|}{$-0.04^{+0.14}_{-0.24}$} & \multicolumn{1}{|c|} {$0.05^{+0.05}_{-0.34}$} & \multicolumn{1}{|c}{$-0.17^{+0.24}_{-0.12}$} \\
\hline
 \end{tabular}
\end{center}
\end{table*}

In Table~\ref{tab:fitting}, we list the MCMC fitting results for Union2.1 and JLA data with and without systematic errors, including both cosmological parameters and light-curve parameters. If we compare our JLA constraints in Table~\ref{tab:fitting} with the results in~table 10 of \cite{Betoule14}, we can see that our constraints of [$\Omega_{\rm M}$, $\alpha$, $\beta$, $M_{B}$] on flat $w$CDM model is consistent with the results in Ref.~\cite{Betoule14} within 1-$\sigma$ CL. Then we compare our Union2.1 results with table 6 and 7 in \cite{Suzuki2012}, and we also find good agreements between these two results. Although the best-fit values of $\Omega_{\rm M}$ and $w$ are larger than theirs for flat $w$CDM model, they are still consistent in 1-$\sigma$. Note that our constraint errors are basically larger than that in \cite{Suzuki2012} and \cite{Betoule14}, and this is due to that we set $h$ as a free parameter in our constraints and thus have one more free parameters than theris.

In our constraint results, the cosmological parameters from Union2.1 and JLA data are consistent with each other in 1-$\sigma$ confidence level for the corresponding models, but the light-curve parameters are a bit different since they are using different parameterization for the correction between $M_B$ and host galaxy mass. In addition, we also show the best-fits and 1-$\sigma$ errors for $h$ in Table \ref{tab:fitting}, but we should keep it in mind that its PDFs are relatively flat and it is not well constrained by SN Ia data. It is similar for $M_B$ and $\Delta_M$ in JLA data, that the PDFs are flat over the whole range, and just small and smooth bumps are shown in the PDF profile.

\subsection{BAO data}

The baryon acoustic oscillation is a good measurement for the cosmic distance and evolution. The BAO scale can be derived by fitting the scale dilation factor $\alpha$, which depends on $D_{\rm V}/r_{\rm d}$, according to a fiducial cosmological model. The $D_{\rm V}$ is the volume weighted effective distance at redshift $z$, which can be expressed as
\be
D_{\rm V}(z)\equiv \left[(1+z)^{2}D^{2}_{\rm A}(z)\frac{cz}{H(z)} \right]^{1/3} \label{eq:DV},
\ee
Where $D_{\rm A}(z)=D_{\rm C}(z)/(1+z)$ is the angular diameter distance, and we can also define a radial scale $D_{\rm H}(z)\equiv c/H(z)$. The $D_{\rm A}$ and $D_{\rm H}$ denote the distances can be measured by the BAO modes that perpendicular to and along the line of sight, respectively.

We use several BAO galaxy clustering observations from six-degree-field galaxy survey (6dFGS) \cite{Beutler11}, Sloan Digital Sky Survey Data Release 7 and 11(SDSS DR7 and DR11) \cite{Ross14,Anderson14}, WiggleZ dark energy survey \cite{Kazin14}, and the Ly$\alpha$ forest measurements from Baryon Oscillation Spectroscopic Data Release 11 (BOSS DR11) \cite{Delubac15,Font14}. The $\rsdv(z)$ values derived from these surveys are listed in Table~\ref{tab:BAO_data}.

\begin{table}[!t]
\caption{The BAO data used in this work.} \label{tab:BAO_data}
\vspace{1mm}
\begin{center}
\begin{tabular}{c|c|c}
\hline  \hline
Redshift & $\rsdv(z)$ & Data set\\
\hline
0.1 & 0.336$\pm$0.015 & 6dF \cite{Beutler11}\\
0.15 & 0.2239$\pm$0.0084$^a$ & SDSS DR7 \cite{Ross14}\\
0.32 & 0.1181$\pm$0.0023$^a$ & SDSS-III DR11 \cite{Anderson14}\\
0.57 & 0.0726$\pm$0.0007$^a$ & SDSS-III DR11 \cite{Anderson14}\\
0.44 & 0.0870$\pm$0.0042 & WiggleZ \cite{Kazin14}\\
0.60 & 0.0672$\pm$0.0031 & WiggleZ \cite{Kazin14}\\
0.73 & 0.0593$\pm$0.0020 & WiggleZ \cite{Kazin14}\\
2.34 & 0.0320$\pm$0.0013$^b$ & SDSS-III DR11 \cite{Delubac15}\\
2.36 & 0.0329$\pm$0.0009$^b$ & SDSS-III DR11 \cite{Font14}\\
\hline
\end{tabular}
\end{center}
\vspace{-2mm}
$^a$ The SDSS values here have been inverted from the published values of $D_{\rm V}(z)/r_{\rm d}$, see the details in the text.\\
$^b$ The BOSS values here are estimated from $D_{\rm A}(z)/r_{\rm d}$ and $D_{\rm H}(z)/r_{\rm d}$ in the relevant references, and the details of estimation can be found in the text.\\
\end{table}

The 6dFGS sample is obtained from more than $70,000$ half-sky samples out to $z=0.15$ with effective redshift $z_{\rm eff}=0.106$. The BAO signal is detected at 105 $h^{-1}$Mpc. The SDSS DR7 Main Galaxy Samples (MGS) we use contain 63,163 galaxies and cover 6813 deg$^2$ at $z<0.2$, and the survey gives $D_{\rm V}(z_{\rm eff} = 0.15) = (664 \pm 25)(r_{\rm d}/r_{\rm d,fid})$ Mpc~\cite{Ross14}. The SDSS-III DR11 samples include nearly one million galaxies and cover approximately 8500 deg$^2$ at $0.2<z<0.7$, and they give $D_{\rm V}(z=0.32)=(1264 \pm 25 {\rm Mpc})(r_{\rm d}/r_{\rm d,fid})$ and $D_{\rm V}(z=0.57)=(2056 \pm 20 {\rm Mpc})(r_{\rm d}/r_{\rm d,fid})$\footnote{This sample is the most precise BAO constraint ever obtained from galaxy survey.}, where $r_{\rm d, fid}=149.28$~Mpc \cite{Anderson14}. To unify the measured BAO quantity as $r_{\rm d}/D_{\rm V}(z)$, we make conversion for the results of SDSS DR7 and DR11 and show them in Table \ref{tab:BAO_data}. The WiggleZ Dark Energy Survey samples contain about $200,000$ redshifts of UV-selected galaxies, covering of order $1000$ deg$^{2}$ of equatorial sky, and provides the BAO measurements at three redshifts \cite{Kazin14}. In~\cite{Kazin14}, it is shown the $3 \times 3$ covariance matrix of $D_{\rm V}(r^{\rm fid}_{\rm s}/r_{\rm s})$ for the three redshift points, and we convert this to the $3 \times 3$ covariance matrix of $(r_{\rm d}/D_{\rm V})$ as
\begin{eqnarray}
C=\left(
\begin{array}{ccc}
17.72 & 6.9271 & 0 \\
6.9217 & 9.2720 & 2.2243 \\
0 & 2.2243 & 4.1173
\end{array}%
\right) \times 10^{-6}. \label{eq:C-wiggle}
\end{eqnarray}
Thus, the quoted errors of three WiggleZ data in Table~\ref{tab:BAO_data} are the square roots of the the diagonal values of $C$ matrix. The two correlation coefficients are $\rho=0.54$ and $0.36$ respectively, and the 1$^{\rm st}$ and 3$^{\rm rd}$ redshift bins are uncorrelated.

Beside the galaxy surveys, we also include the measurements of the Ly$\alpha$ forest results from DR11 of BOSS from SDSS-III at redshifts as high as $z=2.34$ and $2.36$ \cite{Delubac15,Font14}. In Ref~\cite{Delubac15}, the flux auto-correlation of Ly$\alpha$ forest is used to detect the BAO features. There are $137,526$ quasars in redshift range $2.1<z<3.5$ from DR11 of BOSS from SDSS-III are included to determine the position of BAO peak, and they find $D_{\rm A}(z=2.34)/r_{\rm d}=11.28\pm0.65$ and $D_{\rm H}(z=2.34)/r_{\rm d}=9.18\pm0.28$. In Ref.~\cite{Font14}, the cross-correlation of quasars with the Ly$\alpha$ forest absorption is measured. The BAO scale is then given by $D_{\rm A}(z=2.36)/r_{\rm d}=10.8\pm0.4$, and $D_{\rm H}(z=2.36)/r_{\rm d}=9.0\pm0.3$. Then, we use Eq.(\ref{eq:DV}) to calculate $\rsdv$ and estimate the errors from the values and errors of $D_{\rm A}/r_{\rm d}$ and $D_{\rm H}/r_{\rm d}$.

\subsection{CMB data}

Before the epoch of recombination, the baryons are coupled with photons by the large density of free electrons, and the photon-baryon fluid propagates as acoustic waves. We can define the comoving sound horizon of this wave that can reach before the baryon decoupling or the end of baryon drag epoch. This sound horizon depends only on the physics of the early Universe, which is estimated by
\ba
r_{\rm d}(z_{\rm d})&=& \int^{t_{\rm s}}_{0} c_{\rm s} \frac{\der t}{a} \nonumber \\
&=&
\frac{c}{\sqrt{3}}\int^{a_{\rm d}}_{0}\frac{\der a}{a^{2}H(a)(1+(3\Omega_{\rm b}/4\Omega_{\gamma})a)^{1/2}},
\ea
where $z_{\rm d}$ is the redshift of the drag epoch, $c_{\rm s}$ is sound speed, $t_s$ denotes the epoch of last scattering, $a$ is the scalar factor and $a_{\rm d}=1/(1+z_{\rm d})$, and $\Omega_{\rm b}$ and $\Omega_{\gamma}$ are the fractional energy densities of baryon and radiation, respectively.

The most state-of-the-art CMB observational data is the data from {\it Planck} 2015 results. Here we directly use the constraint on $r_{\rm d}$, i.e. the comoving sound horizon at the drag epoch, from the {\it Planck} 2015 results from cosmological parameters \cite{Planck15-13}
\begin{eqnarray}
r_{\rm d}/{\rm Mpc}=147.27 \pm 0.31 \text{  }(1\sigma\text{  }{\rm CL}). \label{eq:rd}
\end{eqnarray}
This value is derived from the combined constraint of {\it Planck} TT, TE, EE power spectrum~\cite{Planck15-13}. We note that the $r_{\rm d}$ is almost the same for $\rm \Lambda$CDM and $w$CDM models and the discrepancy is less than 0.15\% \cite{Hinshaw13,Planck15-13}, since the dark energy is not important to affect the evolution of the Universe at early time. Therefore, we use the same value of $r_{\rm d}$ given by Eq. (\ref{eq:rd}) for both $\rm \Lambda$CDM and $w$CDM models in our following estimation.

\section{Comparison}
\label{sec:compare}

\begin{table*}
\begin{center}
\caption{The $\chi^2$ and PTE for Union2.1, JLA data sets with the results of $Planck$ 15 (P15, flat $\Lambda$CDM) as hypothetic model. For comparison, we also show the minimum $\chi^2$ and the corresponding PTE for non-flat $\Lambda$CDM and flat $w$CDM models from MCMC fitting for Union2.1, JLA data sets. The degree of freedoms are 572 and 732 for Union2.1 and JLA data, respectively.}  \label{tab:PTE}
\vspace{4mm}
\begin{tabular}{ l | l | l | l | l | l | l | | | | | | | | | l | l | l | l | l | l | l | l | l | l | l | l | l | | |}
\hline\hline
\multicolumn{1}{c|} {Data} & \multicolumn{6}{|c|}{Union2.1}  & \multicolumn{6}{|c}{JLA} \\
\hline 
\multicolumn{1}{c|} {Model} & \multicolumn{2}{|c|} {non-flat $\Lambda$CDM} & \multicolumn{2}{|c|}{flat $w$CDM} & \multicolumn{2}{|c|}{P15 (flat $\Lambda$CDM)}&  \multicolumn{2}{|c|} {non-flat $\Lambda$CDM} & \multicolumn{2}{|c}{flat $w$CDM} & \multicolumn{2}{|c}{P15 (flat $\Lambda$CDM)}\\
 \hline
\multicolumn{1}{c|} {Error} & \multicolumn{1}{|c|}{no sys} & \multicolumn{1}{|c|}{sys} &\multicolumn{1}{|c|}{no sys} & \multicolumn{1}{|c|}{sys} & \multicolumn{1}{|c|}{no sys} & \multicolumn{1}{|c|}{sys} &\multicolumn{1}{|c|}{no sys} & \multicolumn{1}{|c}{sys}  &\multicolumn{1}{|c|}{no sys} & \multicolumn{1}{|c|}{sys} & \multicolumn{1}{|c|}{no sys} & \multicolumn{1}{|c}{sys}\\
\hline
\multicolumn{1}{c|} {$\chi^2$ } & \multicolumn{1}{|c|} {548.89}  & \multicolumn{1}{|c|} {530.33}& \multicolumn{1}{|c|} {548.87}& \multicolumn{1}{|c|} {530.37}& \multicolumn{1}{|c|} {551.76}& \multicolumn{1}{|c|} {530.68}& \multicolumn{1}{|c|} {724.70}& \multicolumn{1}{|c|} {508.30}& \multicolumn{1}{|c|} {724.79}& \multicolumn{1}{|c|} {508.37}& \multicolumn{1}{|c|} {732.98}& \multicolumn{1}{|c} {512.06}\\

\multicolumn{1}{c|} {PTE (\%)} & \multicolumn{1}{|c|} {74.94}  & \multicolumn{1}{|c|} {89.31}& \multicolumn{1}{|c|} {74.96}& \multicolumn{1}{|c|} {89.29}& \multicolumn{1}{|c|} {72.10}& \multicolumn{1}{|c|} {89.11}& \multicolumn{1}{|c|} {56.90}& \multicolumn{1}{|c|} {99.99}& \multicolumn{1}{|c|} {56.81}& \multicolumn{1}{|c|} {99.99}& \multicolumn{1}{|c|} {48.28}& \multicolumn{1}{|c} {99.99}\\
\hline
 \end{tabular}
\end{center}
\end{table*}

\begin{figure}
\centerline{\includegraphics[width=3.3in]{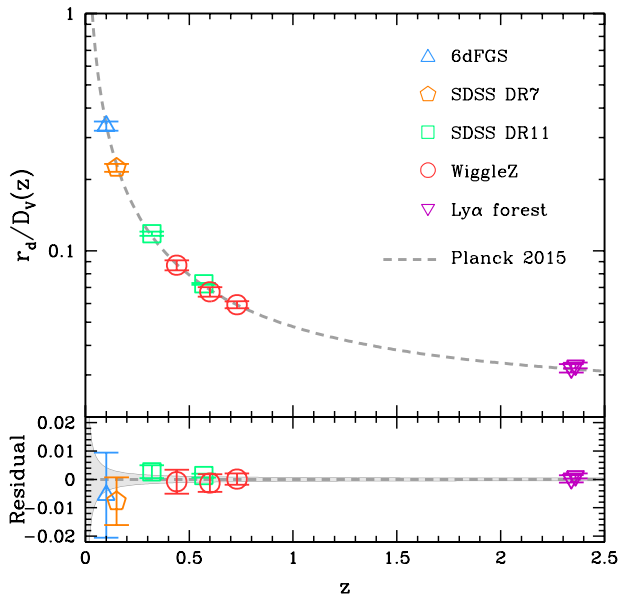}}
\caption{The comparison of $\rsdv(z)$ data from BAO surveys with the prediction from $Planck$ 2015 best-fitting cosmological parameters and 1-$\sigma$ errors (gray dashed line and region). The residual corresponding to the prediction of $Planck$ 2015 is also shown.} \label{fig:rdDv_Planck_BAO}
\end{figure}

\begin{figure*}
\centerline{\includegraphics[width=6.6in]{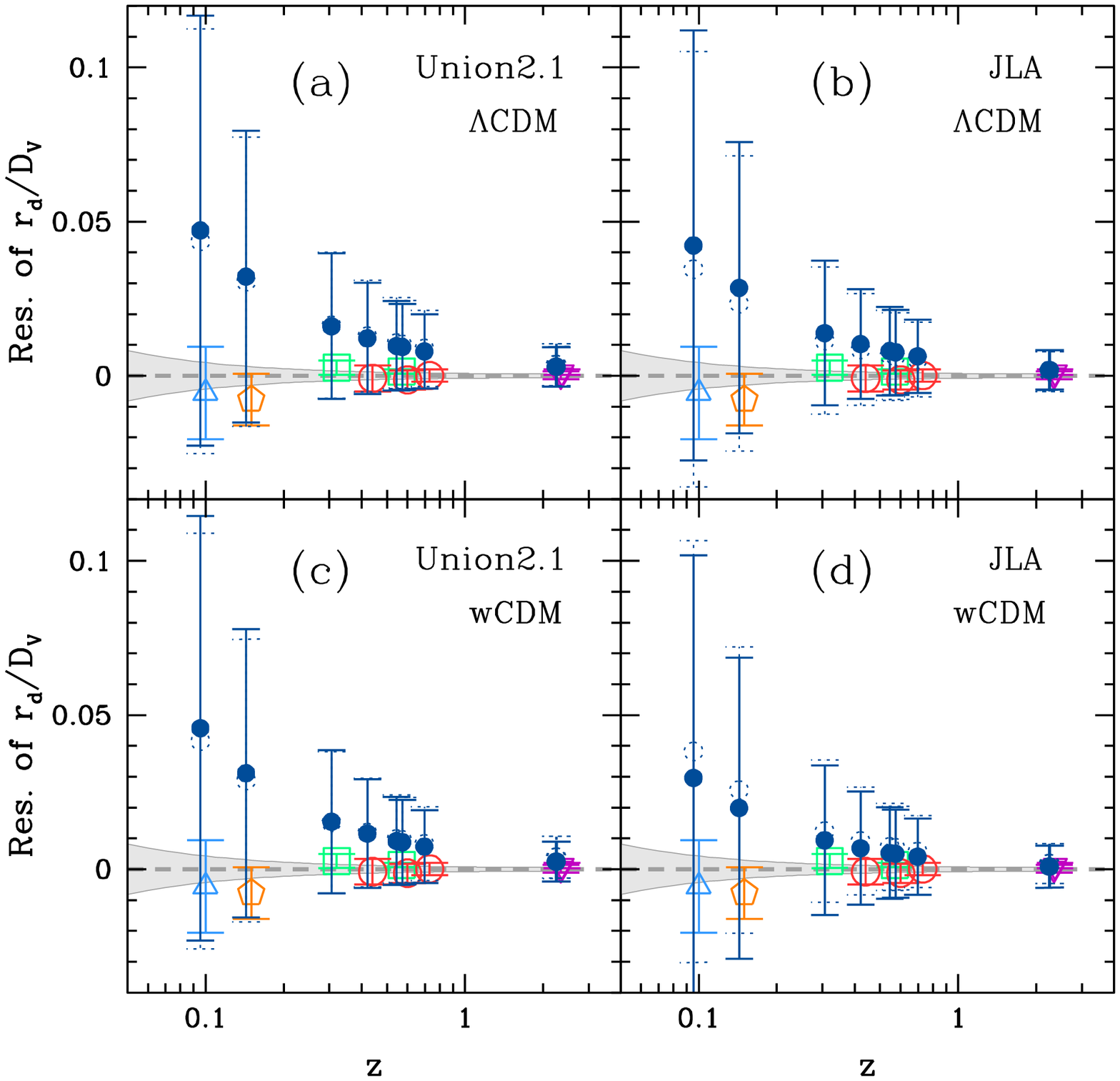}}
\caption{Residuals of $\rsdv(z)$ by subtracting the predictions of best-fitting $Planck$ 2015 results. The dark blue solid (dashed open) points with error bars denote $r_{\rm d}^{\rm CMB}/D_{\rm V}^{\rm SN}(z)$ from $Planck$ 2015 and SN Ia surveys without (with) systematic error. The BAO measurements are also shown with the same symbols as in Fig.~\ref{fig:rdDv_Planck_BAO}. In order to make comparison with the BAO data clearly, we slightly shift the dark blue solid dots towards lower redshifts. Note that the results of $r_{\rm d}^{\rm CMB}/D_{\rm V}^{\rm SN}(z)$ at $z=2.34$ and $2.36$ are very similar with each other, and the two dark blue points with error bars are almost overlapped together at these two redshifts. Panel (a): $D_{\rm V}^{\rm SN}$ derived from Union2.1 data for non-flat $\rm \Lambda$CDM model; Panel (b): $D_{\rm V}^{\rm SN}$ derived from Union2.1 data for flat $w$CDM model; Panel (c): $D_{\rm V}^{\rm SN}$ derived from JLA data for non-flat $\rm \Lambda$CDM model; Panel (d): $D_{\rm V}^{\rm SN}$ derived from JLA data for flat $w$CDM model. We find the $r_{\rm d}^{\rm CMB}/D_{\rm V}^{\rm SN}(z)$ from CMB and SN Ia observations is consistent with $\left[r_{\rm d}/D_{\rm V}(z)\right]^{\rm BAO}$ from BAO measurements in 1$\sigma$ CL.
} \label{fig:residual_rdDv}
\end{figure*}

In this section, we compare and check the consistency between the results of the SN Ia, BAO and CMB measurements. We first check the consistency between the SN data and CMB measurements from $Planck$ 2015 by comparing their $\chi^2$ and probability to exceed (PTE). Next, we compare the BAO results $(\rsdv)^{\rm BAO}$ with CMB results $(\rsdv)^{\rm CMB}$. At last, we perform the comparison using the quantity $r_{\rm d}^{\rm CMB}/D_{\rm V}^{\rm SN}$ with $(\rsdv)^{\rm BAO}$ and $(\rsdv)^{\rm CMB}$ for all three kinds of observations.

In Table~\ref{tab:PTE}, we show the results of $\chi^2$ and the PTE using $Planck$ 2015 results (flat $\Lambda$CDM) as hypothetic model for Union2.1 and JLA data sets. The PTE is given by
\be \label{eq:PTE}
{\rm PTE} = \frac{1}{2^{d/2}\Gamma(d/2)}\int_{\chi^2}^{\infty} t^{d/2-1} e^{-t/2} dt,
\ee
where $d$ is the number of degree of freedom, $\chi^2$ is from the model fitting and $\Gamma(x)$ is the Gamma function.
For comparison, the minimum $\chi^2$ and corresponding PTE from our MCMC fitting for non-flat $\Lambda$CDM and flat $w$CDM models are also shown. The degree of freedoms (dof) are 572 and 732 for Union2.1 and JLA data, respectively (we have seven free parameters in all models). Note that since we have light-curve parameters fitted simultaneously with cosmological parameters in the MCMC process, we use the results of $\Omega_{\rm M}=0.3156$, $\Omega_{\Lambda}=1-\Omega_{\rm M}$ and $w=-1$ from $Planck$ 2015 with the best-fit values of light-curve parameters (using Eq.~(\ref{eq:chi2})) to calculate the corresponding $\chi^2$ and PTE for Union2.1 and JLA data. We find that the $\chi^2$ and PTE of $Planck$ 2015 are comparable with the minimum $\chi^2$ and PTE of non-flat $\Lambda$CDM and flat $w$CDM models for both Union2.1 and JLA data with and without systematic errors. This indicates that the results of $Planck$ 2015 and SN Ia data are in good agreements. Also, the PTEs for all models are large and greater than $0.48$ which implies all these models are good to explain the SN Ia data sets. We note that some of the PTE values are quite high, especially for JLA data with systematic errors, which are close to 1. This implies that the models over-fit the data, and the errors of these data are probably overestimated.

In Fig.~\ref{fig:rdDv_Planck_BAO}, we show the comparison result of the BAO surveys and the predication of the best-fitting cosmological parameters and 1-$\sigma$ errors from $Planck$ 2015. The $\rsdv(z)$ values from the BAO surveys are listed in Table \ref{tab:BAO_data}. The values of cosmological parameters from $Planck$ 2015 we use to calculate $\rsdv(z)$ and errors are $\Omega_{\rm M}=0.3156\pm0.0091$, $\Omega_{\rm \Lambda}=1-\Omega_{\rm M}$ and $H_0=67.27\pm0.66$ $\rm km\,s^{-1}Mpc^{-1}$ \cite{Planck15-13}. As can be seen, the $\rsdv(z)$ of BAO measurements are consistent with the $Planck$ 2015 best-fitting prediction in 1$\sigma$ CL, except for two data of SDSS-III DR11 at $z=0.32$ and $0.57$ and one data point of DR11 of BOSS from SDSS-III at $z=2.36$, whose 1-$\sigma$ lower limits are a bit higher (within 2$\sigma$) than the $Planck$ 2015 result. In order to estimate the combined significance of the consistency, we calculate the effective $\chi^2$ for the BAO data using the best-fitting cosmological parameters from $Planck$ 2015 as model prediction, which is given by 
\begin{eqnarray} \label{eq:chi2_eff}
\chi^2_{\rm eff}&=& \sum^{N}_{i,j} \left[(\rsdv)_{i}^{\rm CMB}-(\rsdv)_{i}^{\rm BAO}\right]\left(C_{\rm BAO}\right)^{-1}_{ij} \nonumber \\
& \times & \left[(\rsdv)_{j}^{\rm CMB}-(\rsdv)_{j}^{\rm BAO}\right],
\end{eqnarray}
where N=9 is the number of BAO data, and $C_{\rm BAO}$ is the full covariance matrix of nine BAO data. For 6dF, SDSS-DR7, SDSS-III DR11, their quoted errors are uncorrelated, so they only have diagonal values in the $C_{\rm BAO}$ matrix (Table~\ref{tab:BAO_data}). But for WiggleZ survey, we use Eq.~(\ref{eq:C-wiggle}) as its covariance matrix. We find $\chi^2_{\rm eff}=7.93$, the reduced $\chi^2_{\rm eff}=0.99$ with dof=8 (we don't have free parameter here),  and the PTE=0.44 with the $Planck$ 2015 results as the hypothetic model. This means the results of $Planck$ 2015 can fit the BAO data very well. Therefore, generally speaking, the two kinds of observations are consistent with each other.

    Next, we perform our more robust test on cosmic evolution by including SN Ia observations. For the numerator of $\rsdv$ (the scale of the sound horizon), since it is determined by the physics before $z\simeq1100$, we only use the CMB constraint from {\it Planck} TT+TE+EE power spectrum which gives $r_{\rm d}=147.27\pm0.31$ Mpc \cite{Planck15-13}. For the denominator of $\rsdv$, i.e. $D_{\rm V}(z)$ which is mostly determined by the low-redshift evolution of the Universe, we adopt the constraints given by supernova measurements from Union2.1 and JLA catalogs. Note that the $D_{\rm V}(z)$ from SN Ia catalogs is derived from our MCMC results. In order to avoid the statistical bias, we actually estimate $1/D_{\rm V}(z)$ for each MCMC chain point, which is composed of the cosmological parameters and light-curve parameters. We then derive the mean values and standard deviations of $1/D_{\rm V}$ at different redshifts. Note that the mean values of  $1/D_{\rm V}(z)$ are obtained by integrating over the full PDFs of $1/D_{\rm V}(z)$ from the MCMC results, which take account of the whole profiles of the PDFs.

\begin{table*}
\begin{center}
\caption{The effective $\chi^2$ for comparisons of the $\rsdv$ results for SN Ia, CMB and BAO data. The reduced $\chi^2_{\rm eff}$ are also shown with dof=8. Besides, we calculate the PTE with $Planck$ 2015 results as hypothetic model for comparing $r_{\rm d}^{\rm CMB}/D_{\rm V}^{\rm SN}$ with $(\rsdv)^{\rm CMB}$.} \label{tab:chi2}
\vspace{1mm}
\begin{tabular}{ l | l | l | l | l | l | l | | | | | | | | | l | l | l | l | l | l | l | l | l | l | l | l | l | | | l | l | l | l | l | | | | | | | | | l | l  | l | l | l | l | | | | | | | | | l | l | l | l | l | l | l | l | l | l | l | l | l | | | l}
\hline\hline
\multicolumn{1}{c|} {Data} & \multicolumn{8}{|c|}{Union2.1}  & \multicolumn{8}{|c}{JLA} \\
\hline 
& \multicolumn{4}{|c|}{$r_{\rm d}^{\rm CMB}/D_{\rm V}^{\rm SN}$ \& $(\rsdv)^{\rm CMB}$}  & \multicolumn{4}{|c|}{$r_{\rm d}^{\rm CMB}/D_{\rm V}^{\rm SN}$ \& $(\rsdv)^{\rm BAO}$} & \multicolumn{4}{|c|}{$r_{\rm d}^{\rm CMB}/D_{\rm V}^{\rm SN}$ \& $(\rsdv)^{\rm CMB}$} & \multicolumn{4}{|c}{$r_{\rm d}^{\rm CMB}/D_{\rm V}^{\rm SN}$ \& $(\rsdv)^{\rm BAO}$}\\
\hline
\multicolumn{1}{c|} {Model} & \multicolumn{2}{|c|} {non-flat $\Lambda$CDM} & \multicolumn{2}{|c|}{flat $w$CDM} &  \multicolumn{2}{|c|} {non-flat $\Lambda$CDM} & \multicolumn{2}{|c}{flat $w$CDM} & \multicolumn{2}{|c|} {non-flat $\Lambda$CDM} & \multicolumn{2}{|c|}{flat $w$CDM} &  \multicolumn{2}{|c|} {non-flat $\Lambda$CDM} & \multicolumn{2}{|c}{flat $w$CDM}\\
 \hline
\multicolumn{1}{c|} {Error} & \multicolumn{1}{|c|}{no sys} & \multicolumn{1}{|c|}{sys} &\multicolumn{1}{|c|}{no sys} & \multicolumn{1}{|c|}{sys} & \multicolumn{1}{|c|}{no sys} & \multicolumn{1}{|c|}{sys} &\multicolumn{1}{|c|}{no sys} & \multicolumn{1}{|c}{sys}  &\multicolumn{1}{|c|}{no sys} & \multicolumn{1}{|c|}{sys} & \multicolumn{1}{|c|}{no sys} & \multicolumn{1}{|c}{sys} &\multicolumn{1}{|c|}{no sys} & \multicolumn{1}{|c|}{sys} & \multicolumn{1}{|c|}{no sys} & \multicolumn{1}{|c}{sys}\\
\hline
\multicolumn{1}{c|} {$\chi^2_{\rm eff}$} & \multicolumn{1}{|c|}{3.576} & \multicolumn{1}{|c|}{3.942} &\multicolumn{1}{|c|}{3.252} & \multicolumn{1}{|c|}{3.736} & \multicolumn{1}{|c|}{3.414} & \multicolumn{1}{|c|}{3.722} &\multicolumn{1}{|c|}{3.142} & \multicolumn{1}{|c}{3.508}  &\multicolumn{1}{|c|}{2.497} & \multicolumn{1}{|c|}{1.633} & \multicolumn{1}{|c|}{1.011} & \multicolumn{1}{|c}{2.121} &\multicolumn{1}{|c|}{2.448} & \multicolumn{1}{|c|}{1.636} & \multicolumn{1}{|c|}{1.074} & \multicolumn{1}{|c}{2.085}\\

\multicolumn{1}{c|} {$\chi^2_{\rm red}$} & \multicolumn{1}{|c|}{0.447} & \multicolumn{1}{|c|}{0.493} &\multicolumn{1}{|c|}{0.406} & \multicolumn{1}{|c|}{0.467} & \multicolumn{1}{|c|}{0.427} & \multicolumn{1}{|c|}{0.465} &\multicolumn{1}{|c|}{0.393} & \multicolumn{1}{|c}{0.439}  &\multicolumn{1}{|c|}{0.312} & \multicolumn{1}{|c|}{0.204} & \multicolumn{1}{|c|}{0.126} & \multicolumn{1}{|c}{0.265} &\multicolumn{1}{|c|}{0.306} & \multicolumn{1}{|c|}{0.205} & \multicolumn{1}{|c|}{0.134} & \multicolumn{1}{|c}{0.261}\\

\multicolumn{1}{c|} {PTE(\%)} & \multicolumn{1}{|c|}{89.32} & \multicolumn{1}{|c|}{86.23} &\multicolumn{1}{|c|}{91.75} & \multicolumn{1}{|c|}{88.01} & \multicolumn{1}{|c|}{$-$} & \multicolumn{1}{|c|}{$-$} &\multicolumn{1}{|c|}{$-$} & \multicolumn{1}{|c}{$-$}  &\multicolumn{1}{|c|}{96.18} & \multicolumn{1}{|c|}{99.02} & \multicolumn{1}{|c|}{99.81} & \multicolumn{1}{|c}{97.70} &\multicolumn{1}{|c|}{$-$} & \multicolumn{1}{|c|}{$-$} & \multicolumn{1}{|c|}{$-$} & \multicolumn{1}{|c}{$-$}\\
\hline
 \end{tabular}
\end{center}
\end{table*}

Then we combine the $r_{\rm d}$ from $Planck$ 2015 and the mean $1/D_{\rm V}(z)$ from SN Ia catalogs to construct the quantity $r_{\rm d}^{\rm CMB}/D_{\rm V}^{\rm SN}(z)$\footnote{Hereafter, for simplicity, we use the notation $1/D_{\rm V}^{\rm SN}$, $1/D_{\rm A}^{\rm SN}$ and $1/D_{\rm H}^{\rm SN}$ to denote the mean values of these quantities from the SN data, which are obtained by integrating over their full PDFs from the MCMC results.}, and compare it to the results of $\left[r_{\rm d}/D_{\rm V}(z)\right]^{\rm CMB}$ from $Planck$ 2015  and BAO measurements $\left[r_{\rm d}/D_{\rm V}(z)\right]^{\rm BAO}$. We show the comparison results in Fig.~\ref{fig:residual_rdDv}. For clarity, the residuals of $\rsdv(z)$ corresponding to the predictions from $Planck$ 2015 best-fitting results are shown. We derive $r_{\rm d}^{\rm CMB}/D_{\rm V}^{\rm SN}(z)$ for non-flat $\rm \Lambda$CDM and flat $w$CDM models for Union2.1 and JLA data with and without systematic errors. We find the $r_{\rm d}^{\rm CMB}/D_{\rm V}^{\rm SN}(z)$ results are consistent with both $\left[r_{\rm d}/D_{\rm V}(z)\right]^{\rm CMB}$ and $\left[r_{\rm d}/D_{\rm V}(z)\right]^{\rm BAO}$ in 1$\sigma$ for all cases we consider, and there is no strong evidence of the derivation of standard $\rm \Lambda$CDM cosmology evolution. The consistency of the non-flat $\rm \Lambda$CDM model is similar to that of flat $w$CDM model. Also, the values of $r_{\rm d}^{\rm CMB}/D_{\rm V}^{\rm SN}(z)$ are generally higher than $\left[r_{\rm d}/D_{\rm V}(z)\right]^{\rm BAO}$, which indicates the $D_{\rm V}(z)$ obtained from the SN Ia data is smaller than that given by the BAO measurements, especially at low redshifts ($z<0.3$).  

In Table~\ref{tab:chi2}, we show the effective $\chi^2$ for $r_{\rm d}^{\rm CMB}/D_{\rm V}^{\rm SN}$ vs. $(\rsdv)^{\rm CMB}$ and $(\rsdv)^{\rm BAO}$ for different models and SN Ia data sets. The $\chi^2_{\rm eff}$ calculations here are similar to Eq.~(\ref{eq:chi2_eff}), when comparing $r_{\rm d}^{\rm CMB}/D_{\rm V}^{\rm SN}$ with $(\rsdv)^{\rm BAO}$ we use $C=C_{\rm BAO}+C^{\rm diag}_{\rm SN,CMB}$, where $C^{\rm diag}_{\rm SN,CMB}=\sigma^2_{\rm SN,CMB}$ is the error of $r_{\rm d}^{\rm CMB}/D_{\rm V}^{\rm SN}$ at each redshift. We find the effective $\chi^2$ is small and less than 4 for all cases, and the reduced $\chi^2_{\rm eff}$ is less than 0.5. This indicates that all of the data sets are consistent with each other very well from the calculations of $\rsdv(z)$. Besides, using Eq.~(\ref{eq:PTE}) and Eq.~({\ref{eq:chi2_eff}}), we calculate the PTE for comparing $r_{\rm d}^{\rm CMB}/D_{\rm V}^{\rm SN}$ with $(\rsdv)^{\rm CMB}$ with the results of $Planck$ 2015 as the hypothetic model. Here, the $\chi^2$ in Eq.~(\ref{eq:PTE}) are derived from the $\chi^2_{\rm eff}$ by Eq.~(\ref{eq:chi2_eff}) with $C=\sigma^2_{\rm CMB}+C^{\rm diag}_{\rm SN,CMB}$ where $\sigma^2_{\rm CMB}$ is the CMB error from the result of $Planck$ 2015. We find the PTE are greater than 0.88 for all cases which means that  $(\rsdv)^{\rm CMB}$ is well consistent with $r_{\rm d}^{\rm CMB}/D_{\rm V}^{\rm SN}$.

The BAO data at $z=2.34$ and 2.36 from the auto-correlation of Ly$\alpha$ forest and cross-correlation of Ly$\alpha$ and quasars are important for the comparison, since they are located at the median redshifts compared to the SN Ia surveys ($z<1.5$) and CMB measurements. To make further comparison for these two data points, we directly use $\rsda(z)$ and $\rsdh(z)$ given by the Ly$\alpha$ forest measurements\footnote{For uniformity, We inverse $\dars$ and $\dhrs$ given by the Ly$\alpha$ forest measurements to derive $\rsda(z=2.34)=0.0887\pm0.0051$ and $\rsdh(z=2.34)=0.1089\pm0.0033$, and $\rsda(z=2.36)=0.0926\pm0.0034$ and $\rsdh(z=2.36)=0.1111\pm0.0037$.}. We then derive the mean values and standard deviations of $1/D_{\rm A}(z)$ and $1/D_{\rm H}(z)$ from our MCMC chain points for SN Ia data, and get the corresponding $r_{\rm d}^{\rm CMB}/D_{\rm A}^{\rm SN}(z)$ and $r_{\rm d}^{\rm CMB}/D_{\rm H}^{\rm SN}(z)$ at $z=2.34$ and 2.36 with $r_{\rm d}^{\rm CMB}=147.27 \pm 0.31$ Mpc given by $Planck$ 2015. 

\begin{figure*}
\centerline{\includegraphics[width=3.2in]{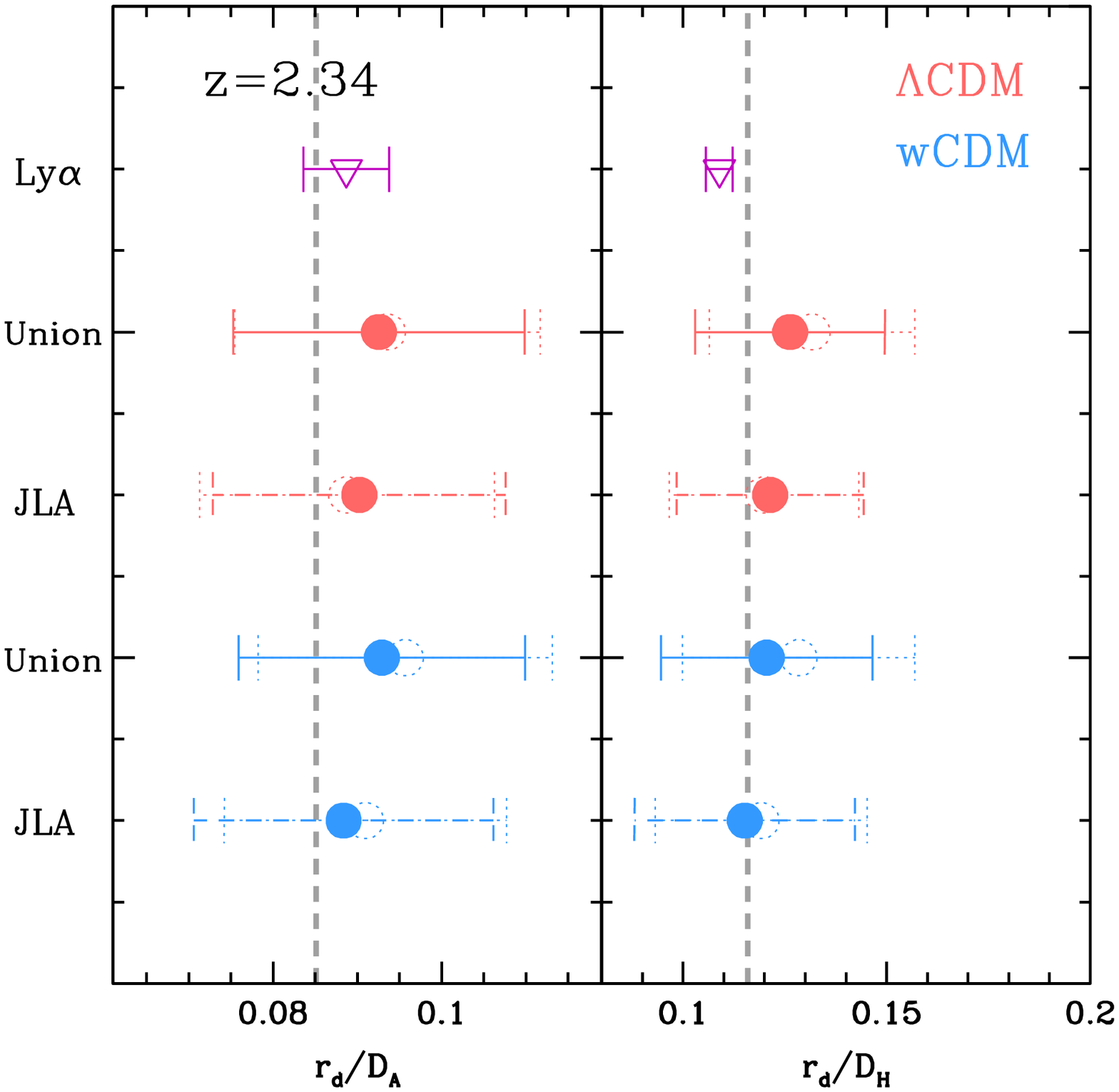}
\includegraphics[width=3.2in]{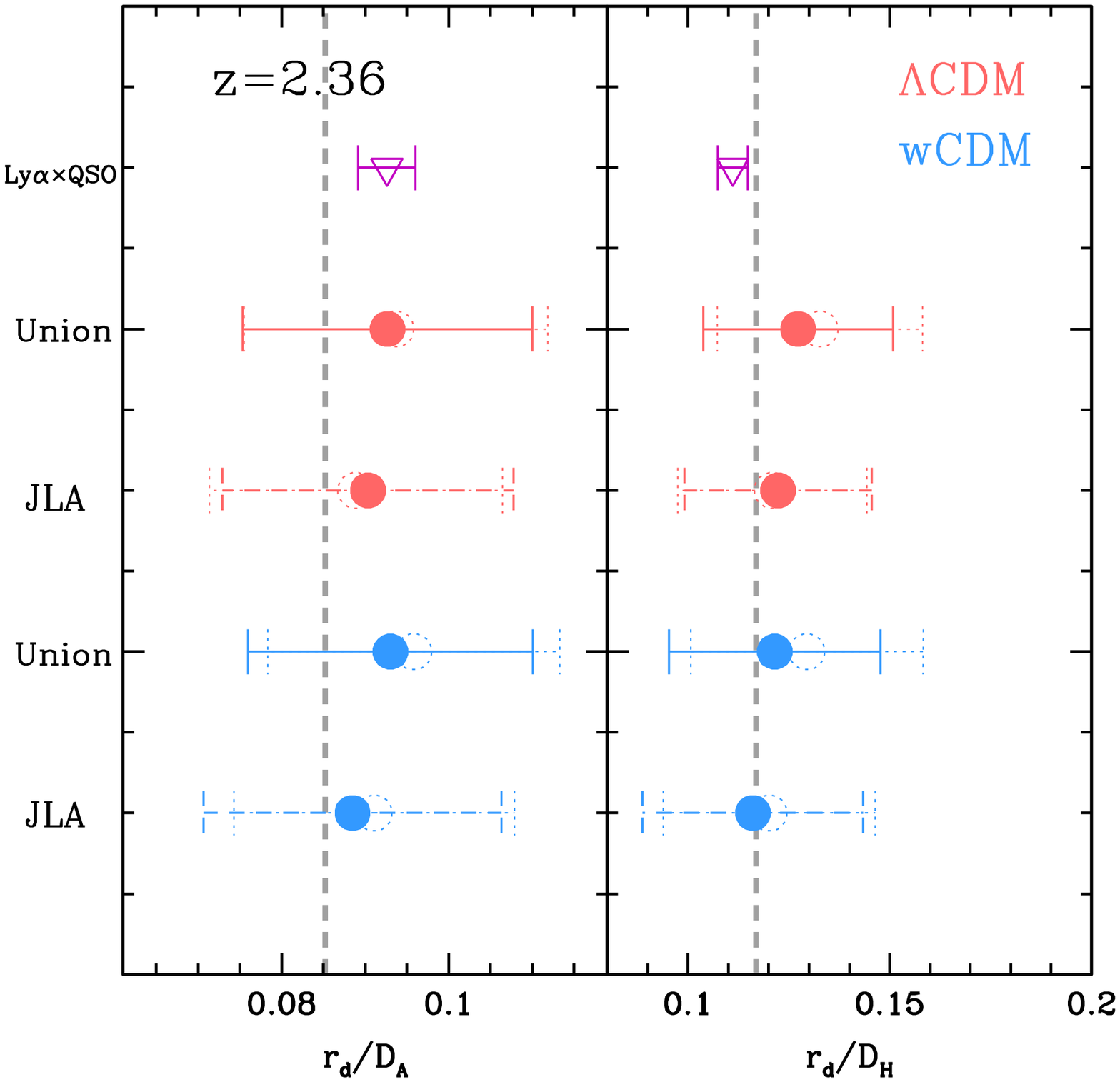}}
\caption{The comparison between the direct measurements of $\rsda$ and $\rsdh$ from Ly$\alpha$ forest and predicted values from SN Ia data sets and $Planck$ 2015 at $z=2.34$ and 2.36 with (dotted open circles) and without (solid points) SN Ia systematic errors. The grey vertical line is the prediction from {\it Planck} 2015 best-fitting cosmological parameters. We find the discrepancies are larger than 1$\sigma$ between Ly$\alpha$ forest (purple inversed triangles) and {\it Planck} 2015 best-fits (gray lines), except for the data $\rsda(z=2.34)$. The consistency of the $\rm \Lambda$CDM model is similar to that of $w$CDM model, and they are consistent with the Ly$\alpha$ forest and CMB results in 1$\sigma$ CL.} \label{fig:rdDADH}
\end{figure*}

In Fig.~\ref{fig:rdDADH}, we show the comparison results of $r_{\rm d}^{\rm CMB}/D_{\rm A}^{\rm SN}(z)$ and $r_{\rm d}^{\rm CMB}/D_{\rm H}^{\rm SN}(z)$ with $\left[r_{\rm d}/D_{\rm A}(z)\right]^{\rm BAO}$ and $\left[r_{\rm d}/D_{\rm H}(z)\right]^{\rm BAO}$ from Ly$\alpha$ BAO measurements, respectively. Both of the cases with and without SN Ia systematic errors are shown in dotted open and solid points with error bars. Similar to Fig.~\ref{fig:residual_rdDv}, we compare the results for both non-flat $\rm \Lambda$CDM and $w$CDM models with Union2.1 and JLA data. We find the discrepancies are larger than 1$\sigma$ between Ly$\alpha$ forest (purple inversed triangles) and {\it Planck} 2015 best-fits (gray lines), except for the data $\rsda(z=2.34)$. The discrepancies shown here are clearer than the $\rsdv$ results shown in Fig.~\ref{fig:rdDv_Planck_BAO}, since the $\rsdv$ is derived by combining $\rsda$ and $\rsdh$. For the comparison including the SN Ia data, the results at $z=2.34$ and 2.36 are quite similar with each other. The consistency of the $\rm \Lambda$CDM model is similar to that of $w$CDM model, and they are consistent with all cases of the Ly$\alpha$ forest measurements and CMB results in 1$\sigma$ CL. We don't find significant deviation between SN Ia, BAO and CMB data.

\section{Summary and conclusion}
\label{sec:conclude}

In this paper, we propose a robust method of testing the consistency of the cosmic evolution for a given cosmological model. Our spirit is that if a model can fully describe the the cosmic evolution throughout the history of the Universe, it should be able to fit the observational data from the high redshifts to low redshifts. We find that the measurement of BAO is a sensitive indicator for this comparison, since it depends on both $r_{\rm d}$, which is determined from physics before recombination, and $D_{\rm V}$ which relies on the evolution of the low-redshift Universe. Then we can compare $\rsdv$ from BAO with the $r_{\rm d}$ obtained from the CMB observations and the $D_{\rm V}$ derived from the low-redshift SNe Ia measurements, and check the consistency for these observations given a cosmological model.

We use the SN Ia data from Union2.1 and JLA data sets, which are currently the two largest SN Ia data sets. In order to get the $D_{\rm V}$ and avoid bias on the constraint results, we free all parameters in the $\rm \Lambda$CDM and $w$CDM models, especially for $H_0$, as well as the parameters of the light-curve. In the fitting process, we adopt MCMC technique and illustrate the probability distribution function for each free parameter, and derive the mean values and standard deviations of $1/D_{\rm V}^{\rm SN}$ at different redshifts. We use the BAO measurements from 6dfGS, SDSS DR7, SDSS DR11, WiggleZ and DR11 of SDSS from SDSS-III that the highest redshift can reach $z=2.36$, and unify the BAO results to be $\left[\rsdv\right]^{\rm BAO}$ for all BAO data. The $r_{\rm d}^{\rm CMB}$ is taken from $Planck$ 2015 results, and we use the same $r_{\rm d}^{\rm CMB}$ for both $\rm \Lambda$CDM and $w$CDM models, since it is quite similar for different dark energy models.

We find that the observed BAO value is well consistent with the derived quantity of $r_{\rm d}^{\rm CMB}/D_{\rm V}^{\rm SN}$ over the redshifts from 0.1 to 2.36 for both $\rm \Lambda$CDM and $w$CDM models. The consistency of non-flat $\rm \Lambda$CDM model is similar to the flat $w$CDM model for all cases we consider. We also note that the $r_{\rm d}^{\rm CMB}/D_{\rm V}^{\rm SN}$ at $z<0.3$ is higher than the $\left[\rsdv\right]^{\rm BAO}$ for both $\rm \Lambda$CDM and $w$CDM models. This is probably because that the derived $D_{\rm V}^{\rm SN}$ tends to be smaller than that from BAO at low redshifts. In order to further check the consistency of the BAO data at median redshifts ($z=2.34$ and $2.36$) provided by the Ly$\alpha$ forest measurements, we also compare $\left[\rsda\right]^{\rm BAO}$ for transverse direction and $\left[\rsdh\right]^{\rm BAO}$ for radial direction with $r_{\rm d}^{\rm CMB}/D_{\rm A}^{\rm SN}$ and $r_{\rm d}^{\rm CMB}/D_{\rm H}^{\rm SN}$, respectively. We find that the $r_{\rm d}^{\rm CMB}/D_{\rm A}^{\rm SN}$ for both of $\rm \Lambda$CDM and $w$CDM models are in good agreements with $\left[\rsda\right]^{\rm BAO}$ measurements in 1-$\sigma$ CL. at $z=2.34$ and $2.36$. These results indicate that the SN Ia, BAO and CMB data are in good agreements, and there is no significant deviation from the standard $\Lambda$CDM model. Also, we should note that the uncertainties of the SN Ia data are probably overestimated in some cases as shown by the PTE values (e.g. see Table \ref{tab:PTE}). This implies that there can be potential deviations from the standard $\Lambda$CDM model hidden in the overestimated uncertainties, which need further confirmation by more accurate measurements from future SN surveys.

In the future, the observations of 21-cm intensity mapping will provide more BAO data across all different redshifts. The BINGO telescope (BAO as Integrated Neutral Gas Observation)~\cite{Battye12,Dickinson14} will cover $0.13<z<0.48$,  CHIME survey (Canadian Hydrogen Intensity Mapping Experiment)~\cite{Newburgh14} will cover $0.8<z<2.5$, and future SKA (Square Kilometer Array) phase-1 will cover cosmic evolution up to redshift 2. The supernovae surveys, such as LSST (Large Synoptic Survey Telescope) which would provide a few hundred thousand or more SNe per year \cite{LSST2009}, could greatly enhance the measurements of the evolution of the low-redshift Universe. All of these future observations will offer more accurate and reliable measurements on the cosmic evolution, and provide stricter constraints on the cosmological models.

\textit{Acknowledgements}--
YG acknowledges the support of Bairen program from the National Astronomical Observatories, Chinese Academy of Sciences. YZM acknowledges support from an ERC Starting Grant (no. 307209). SNZ acknowledges partial funding support by 973 Program of China under grant 2014CB845802, and by the National Natural Science Foundation of China under grants 11373036 and 11133002, the Qianren start-up grant 292012312D1117210, and by the Strategic Priority Research Program "The Emergence of Cosmological Structures" of the Chinese Academy of Sciences, Grant No. XDB09000000. XLC acknowledges the support of the MoST 863 program grant 2012AA121701, pilot B grant XDB09020301, and the NSFC grants 11373030.

\end{document}